\documentclass{article}

\usepackage[dblblindworkshop, final]{neurips_2026}
\workshoptitle{2nd Workshop on Agentic AI Benchmarks and Applications for Enterprise Tasks}

\usepackage[utf8]{inputenc}
\usepackage[T1]{fontenc}
\usepackage{amsmath}
\usepackage{amssymb}
\usepackage{array}
\usepackage{booktabs}
\usepackage{float}
\usepackage{listings}
\usepackage{lineno} 
\usepackage{microtype}
\usepackage{multirow}
\usepackage{tabularx}
\usepackage{xcolor}
\usepackage{url}

\definecolor{linkcol}{HTML}{1F6E9C}
\usepackage[colorlinks=true,linkcolor=linkcol,citecolor=linkcol,
            urlcolor=linkcol,breaklinks=true]{hyperref}

\newcommand{\fico}{\textsc{FiCo}}
\newcolumntype{Y}{>{\raggedright\arraybackslash}X}

\begin{document}

\title{\fico: Find-Then-Compute for Corpus-Level Spreadsheet Question Answering}
\author{%
  Sandarsita Guntupalli \\
  Carnegie Mellon University \\
  \texttt{sandarsg@andrew.cmu.edu} \\
  \And
  Lu He \\
  Atlassian \\
  \texttt{lhe@atlassian.com} \\
  \And
  Kang Li \\
  Atlassian \\
  \texttt{kli3@atlassian.com}
}
\maketitle

\begin{abstract}
Question answering over spreadsheet collections requires finding the correct workbook and computing over complete tables. We introduce Find-then-Compute (\fico{}), which retrieves document summaries, disambiguates similar workbooks, and executes constrained Structured Query Language (SQL) over the selected full table. On DataBench (80 datasets, 1{,}810 questions), \fico{} reaches 76.2\% accuracy: 9.9 points above a strong TableRAG-style baseline on the same frozen workbook choices (66.3\%) under the tracks' prespecified evaluators, and 63.4 points above prefix RAG (12.8\%). On 508 MiMoTable questions, \fico{} reaches 79.7\%, versus 22.2\% for prefix RAG. Giving the strong baseline the gold workbook raises it from 66.3\% to 76.3\%, exposing a 10.0-point source-selection cost under fixed compute. Despite 95.1\% document recall and 98.5\% executable SQL, only 81.3\% of questions execute on the gold workbook. FiCo's advantage comes from integrating semantic source selection with exact, schema-grounded computation.
\end{abstract}

\section{Introduction}

Retrieval-augmented generation (RAG) works when an answer is stated in a retrievable passage~\citep{lewis2020retrieval}. Spreadsheet questions often instead require a count, comparison, or group-by over many rows. A bounded textual context may contain the right schema while omitting values that change the answer. For example, answering ``How many billionaires are under 40?'' requires counting all qualifying rows in a 2{,}668-row table. A three-row prefix cannot determine the answer of 87.

A collection adds a separate source-selection problem. The system must identify the correct workbook and sheet before it can compute. Our operational reference is an engineering reproduction of a deployed prefix-RAG configuration. It serializes inferred headers and at most three rows per sheet, truncates each representation to 2{,}000 tokens, retrieves by embedding similarity, and asks a language model to answer from the retrieved text. The system may find the right workbook yet never expose the rows needed for computation.

We make three contributions. First, we formulate \emph{corpus-level spreadsheet question answering}, where source selection is part of the task, and show that four textual representations reach at most 39\% evidence coverage under a fixed top-5 budget on an aggregation-heavy diagnostic set. Second, we introduce \fico{}, which combines document-summary retrieval, schema-aware disambiguation, and bounded whole-table execution. Third, we compare \fico{} with both prefix RAG and a strong TableRAG-style baseline, then use a paired gold-workbook intervention and component diagnostics to explain the remaining gap between retrieval, execution, and answer correctness.

\paragraph{Relation to prior work.}
Table question-answering and natural-language database-query benchmarks usually provide the relevant table or database~\citep{pasupat2015compositional,yu2018spider,li2023bird,oses2024databench}. SpreadsheetLLM compresses workbook structure~\citep{tian2024spreadsheetllm}, T-RAG retrieves from table corpora~\citep{pan2022trag,zou2025trag}, and FRTR and BRTR retrieve from multimodal workbooks~\citep{gulati2026frtr,gulati2026brtr}. Most closely related, TableRAG couples retrieval with SQL execution for heterogeneous text--table reasoning~\citep{yu2025tablerag}. \fico{} does not claim text-to-SQL novelty. Its contribution is the end-to-end setting where the workbook and sheet are unknown, near-duplicate candidates must be disambiguated, and full-table computation must remain grounded in the selected source.

\section{Task and Method}
\label{sec:method}

Given a question $q$ and collection $\mathcal{D}$, \fico{} retrieves candidates, selects a workbook and table, generates an operation from schema context, executes it, and normalizes the result. This decomposition matters because the gold workbook can enter the candidate set yet be mis-selected, an executable query can use the wrong column, and a correct database result can be lost during normalization.

\begin{figure}[!h]
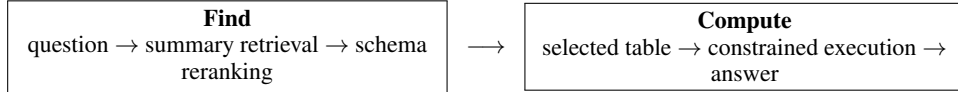

\centering
\small
\begin{tabular}{c@{\hspace{7pt}$\longrightarrow$\hspace{7pt}}c}
\fbox{\parbox{0.40\linewidth}{\centering\textbf{Find}\\question $\rightarrow$ summary retrieval $\rightarrow$ schema reranking}} &
\fbox{\parbox{0.40\linewidth}{\centering\textbf{Compute}\\selected table $\rightarrow$ constrained execution $\rightarrow$ answer}}
\end{tabular}
\caption{\fico{} separates semantic source selection through summary retrieval and schema reranking from exact computation over the selected full table.}
\label{fig:pipeline}
\end{figure}

\paragraph{Why bounded answer-from-text can fail.}
Let $h_q(T)$ be the requested operation on table $T$, and let $C$ be the retrieved context. The answer is not identifiable from $C$ when two tables agree on all information in $C$ but yield different values of $h_q$. This occurs when omitted rows can change the requested statistic and the context contains no sufficient statistic constraining them. The claim concerns answer-from-context under a fixed budget. Retrieval can still locate a table that is then queried exactly.

\paragraph{Find: normalization, summary retrieval, and schema reranking.}
At ingestion time, each source is normalized into a shared SQLite database. Single-table formats (e.g., CSV, TSV) contribute one table, while multi-sheet workbooks (e.g., Google Sheets, Microsoft Excel) contribute one table per sheet. For irregular spreadsheets, we infer the true header row by scoring candidate rows according to column count and cell density. Wide tables that encode years as columns are reshaped into a standard long form. A language model generates a concise summary of each source that describes its topic, tab contents, and column names. We embed one summary per source with \texttt{embeddinggemma-300m}, retrieve the ten most relevant candidates, and rerank them using a \emph{schema signature} of column names and representative values. If the highest-ranked candidate returns an empty result, the system may try the next candidate. This fallback can recover from a wrong initial selection, but can also redirect execution to the wrong source.

\paragraph{Compute: constrained execution with feedback.}
For the selected table, the prompt lists row count, inferred types, numeric ranges or representative values, and up to three sample rows. Gemini~3~Flash generates one read-only SQLite query. Execution errors are returned for at most two repair attempts, for three attempts total. Percentile operations that SQLite does not natively support use a constrained pandas path. The database result is converted to the benchmark answer form. \fico{} is therefore a bounded tool-using agent that selects a source, proposes an operation, observes execution feedback, and may revise the operation.

\section{Experiments}
\label{sec:experiments}

\paragraph{Benchmarks and comparison scope.}
DataBench contributes 1{,}810 questions over 80 datasets and 3{,}708{,}884 rows~\citep{oses2024databench,oses2025semeval}. We also evaluate 508 English questions over 174 multi-sheet MiMoTable workbooks~\citep{li2025mimotable}. In the operational track, the prefix-RAG reproduction and \fico{} share the embedding model, retrieval depth, Gemini~3~Flash generator, and production-style answer judge. Numeric answers use 1\% tolerance, Booleans use exact match, lists use unordered set equality, and ambiguous strings use a shared language-model judge. The baseline isolates the practical change from answer-from-prefix to whole-table execution, but it is not a leaderboard comparison against systems given the correct table.

\paragraph{Controlled component interventions.}
A separately implemented DataBench runner removes one component at a time while holding its loader, model family, prompts, retry policy, synthesis, and judge fixed. Find-stage interventions replace summaries with deterministic schema text, disable schema reranking, or disable empty-result fallback. The compute-stage intervention supplies the gold workbook and removes representative values and sample rows. The benchmark does not provide a gold sheet, so sheet selection remains. We interpret only paired differences within this runner because its absolute accuracy is lower than the operational system.

\paragraph{Strong TableRAG-style baseline.}
We implement a strong retrieval-to-SQL baseline inspired by TableRAG. For a controlled same-source comparison, it receives the workbook choice frozen by the operational \fico{} run; the two systems therefore compute from the same selected source. The baseline uses schema and representative-value context, the benchmark answer type (but not answer), Gemini~3~Flash, two SQL candidates, a typed verifier, up to two feedback-driven repairs, DuckDB execution, and deterministic denotation evaluation. A paired oracle arm changes only the supplied workbook from \fico{}'s selection to the gold workbook. This is a strong TableRAG-style baseline rather than an exact reproduction of TableRAG's heterogeneous text--table implementation.

\paragraph{Fixed-budget retrieval diagnostic.}
On 100 questions over 20 spreadsheets, Coverage@5 records whether five retrieved textual units jointly expose the values or sufficient statistics needed to answer. Four representations reach at most 39\% coverage, while 73 questions require computation. Table~\ref{tab:chunking} gives the results. This is a budget-specific observation, not a universal retrieval ceiling.

\begin{table}[t]
\centering
\footnotesize
\setlength{\tabcolsep}{4.2pt}
\begin{tabular}{lcc}
\toprule
\textbf{Representation} & \textbf{Coverage@5} & \textbf{Answer accuracy} \\
\midrule
350-token chunks & 16\% & 5\% \\
Row packing & 29\% & 10\% \\
Precomputed statistical summaries & 39\% & 15\% \\
Multi-granularity~\citep{gulati2026frtr} & 6\% & 6\% \\
\midrule
Full-table execution reference & --- & \textbf{78\%} \\
\bottomrule
\end{tabular}
\caption{Retrieved text exposes insufficient evidence for most aggregation questions: the best top-5 representation reaches 39\% coverage and 15\% accuracy, versus 78\% for full-table execution. The execution row is a reference, not a matched retrieval ablation.}
\label{tab:chunking}
\end{table}

\section{Results and Analysis}
\label{sec:results}

\begin{table}[!t]
\centering
\footnotesize
\setlength{\tabcolsep}{5pt}
\begin{tabular}{lcc}
\toprule
\multicolumn{3}{l}{\textit{A. Primary accuracy comparison}} \\
\textbf{System} & \textbf{DataBench} & \textbf{MiMoTable} \\
\midrule
Prefix-RAG reproduction & 12.8\% & 22.2\% \\
Strong TableRAG-style baseline & 66.3\% & --- \\
\fico{} & \textbf{76.2\%} & \textbf{79.7\%} \\
$\Delta$ (\fico{} $-$ strong baseline) & $+9.9$ & --- \\
\midrule
\multicolumn{3}{l}{\textit{B. Why the strong baseline falls short}} \\
\textbf{Workbook supplied to baseline} & \textbf{DataBench} & \\
\midrule
Frozen \fico{} selection & 66.3\% & \\
Gold workbook (oracle) & \textbf{76.3\%} & \\
$\Delta$ (gold $-$ selected) & $+10.0$ & \\
\midrule
\multicolumn{3}{l}{\textit{C.} \fico{} \textit{ operational stage cascade}} \\
\textbf{Diagnostic} & \textbf{DataBench} & \textbf{MiMoTable} \\
\midrule
Document recall@10 & 95.1\% & 99.6\% \\
Gold workbook executed & 81.3\% & 90.9\% \\
SQL executes without exception & 98.5\% & 97.2\% \\
Final answer correct & 76.2\% & 79.7\% \\
\bottomrule
\end{tabular}
\caption{\fico{} is 9.9 points above the strong TableRAG-style baseline on the same frozen workbook choices and nearly reaches its 76.3\% gold-workbook oracle. Cross-system values use each track's prespecified evaluator; Panel B alone fixes the compute system and evaluator while changing source identity.}
\label{tab:main}
\end{table}

Table~\ref{tab:main} foregrounds the main result. On the same frozen workbook choices, \fico{} reaches 76.2\% on DataBench, versus 66.3\% for the strong TableRAG-style baseline. The 9.9-point advantage comprises 231 \fico{}-only wins and 52 baseline-only wins; a workbook-clustered bootstrap gives a descriptive 95\% interval of 7.2--12.8 points. Because the two tracks use different prespecified judges, this cross-system gap is descriptive rather than a fully evaluator-matched causal estimate. The paired source intervention below is evaluator-matched.

The operational prefix-RAG reproduction reaches only 12.8\%. \fico{} has 1{,}172 wins and 24 regressions against it; a continuity-corrected McNemar test over the 1{,}196 discordant pairs gives $\chi^2=1100.0$ ($p<10^{-6}$). Thus the result is not merely that SQL execution helps: under fixed source choices, \fico{}'s downstream computation is descriptively stronger than the TableRAG-style baseline, while its complete find--compute pipeline is far stronger than bounded-text RAG.

The stage cascade localizes why this advantage is still imperfect. DataBench document recall is 95.1\%, but the gold workbook is executed on 81.3\% of questions, a 13.8-point loss after retrieval. SQL executes on 98.5\% of questions, yet final accuracy is 76.2\%, a further 22.3-point gap. Executability cannot establish that the source, column, predicate, or aggregation is correct. Gains over prefix RAG span all answer types, from 59.0 points for lists of categories to 68.5 for Booleans.

The gold-workbook arm explains the strong baseline's deficit without making the oracle the paper's focus. Its accuracy rises from 66.3\% to 76.3\% when source identity is supplied, with 190 gold-only wins and 9 selected-only wins (exact McNemar $p=2.9\times10^{-45}$; workbook-clustered 95\% interval, 7.4--12.7 points). For all 1{,}472 questions where \fico{} selected the gold workbook, baseline traces and outcomes are identical across arms, confirming that only source identity changes. The main lesson is that even a strong SQL generator cannot compensate for the wrong workbook; source disambiguation remains a separate, essential part of corpus-level QA.

The component interventions (Appendix~\ref{app:controlled}) support both halves of find-then-compute. Replacing summaries with schema text lowers final accuracy by 2.6 points, and disabling schema reranking lowers it by 0.9. Removing representative values and sample rows leaves execution unchanged but lowers denotation by 7.6 points and final accuracy by 7.0. Thus runnable SQL can remain semantically wrong when the source or value grounding is wrong. A descriptive residual audit similarly assigns 32\% to column confusion and 31\% to near-duplicate workbook selection; these labels lack independent re-annotation and are diagnostic only.

\paragraph{Limitations and deployment considerations.}
Our strong baseline is TableRAG-style rather than an exact reproduction and uses benchmark answer-type metadata. Its strict deterministic evaluator differs from \fico{}'s production-style judge, so the 9.9-point cross-system difference is descriptive; only the baseline's 10.0-point selected-versus-gold intervention is exactly evaluator-matched. The prefix-RAG system is likewise an operational reproduction. Collections of 80 and 174 workbooks do not test larger-scale retrieval, results are single-run, and operational string judgments lack reported human agreement. Cross-sheet joins, formulas, charts, and editing remain unsupported. Deployment also requires latency and cost measurements, access controls, auditable traces, robust SQL parsing, timeouts, row limits, and prompt-injection defenses.

\section{Conclusion}

\fico{} reaches 76.2\% on DataBench, 9.9 points above the strong TableRAG-style baseline and 63.4 above prefix RAG; on MiMoTable it reaches 79.7\%, versus 22.2\% for prefix RAG. Giving the baseline the gold workbook raises it from 66.3\% to 76.3\%, exposing a 10.0-point source-selection cost. FiCo's principal contribution is an integrated agent that finds, disambiguates, and computes over enterprise spreadsheets end to end.

\clearpage
\bibliographystyle{plainnat}
\bibliography{references}

@inproceedings{lewis2020retrieval,
  title     = {Retrieval-Augmented Generation for Knowledge-Intensive {NLP} Tasks},
  author    = {Lewis, Patrick and Perez, Ethan and Piktus, Aleksandra and Petroni, Fabio and Karpukhin, Vladimir and Goyal, Naman and K{\"u}ttler, Heinrich and Lewis, Mike and Yih, Wen-tau and Rockt{\"a}schel, Tim and Riedel, Sebastian and Kiela, Douwe},
  booktitle = {Advances in Neural Information Processing Systems},
  year      = {2020},
  url       = {https://arxiv.org/abs/2005.11401}
}

@inproceedings{pasupat2015compositional,
  title     = {Compositional Semantic Parsing on Semi-Structured Tables},
  author    = {Pasupat, Panupong and Liang, Percy},
  booktitle = {Proceedings of the 53rd Annual Meeting of the Association for Computational Linguistics},
  pages     = {1470--1480},
  year      = {2015},
  url       = {https://aclanthology.org/P15-1142/}
}

@inproceedings{yu2018spider,
  title     = {{Spider}: A Large-Scale Human-Labeled Dataset for Complex and Cross-Domain Semantic Parsing and Text-to-{SQL}},
  author    = {Yu, Tao and Zhang, Rui and Yang, Kai and Yasunaga, Michihiro and Wang, Dongxu and Li, Zifan and Ma, James and Li, Irene and Yao, Qifan and Roman, Shanelle and Zhang, Zilin and Radev, Dragomir},
  booktitle = {Proceedings of the 2018 Conference on Empirical Methods in Natural Language Processing},
  pages     = {3911--3921},
  year      = {2018},
  url       = {https://aclanthology.org/D18-1425/}
}

@inproceedings{li2023bird,
  title     = {Can {LLM} Already Serve as a Database Interface? A Big Bench for Large-Scale Database Grounded Text-to-{SQL}s},
  author    = {Li, Jinyang and Hui, Binyuan and Qu, Ge and Yang, Jiaxi and Li, Binhua and Li, Bowen and Wang, Bailin and Qin, Bowen and Cao, Rongyu and Geng, Ruiying and Huo, Nan and Zhou, Xuanhe and Ma, Chenhao and Li, Guoliang and Chang, Kevin C. C. and Huang, Fei and Cheng, Reynold and Li, Yongbin},
  booktitle = {Advances in Neural Information Processing Systems},
  year      = {2023},
  url       = {https://arxiv.org/abs/2305.03111}
}

@inproceedings{oses2024databench,
  title     = {Question Answering over Tabular Data with {DataBench}: A Large-Scale Empirical Evaluation of {LLM}s},
  author    = {Os{\'e}s Grijalba, Jorge and Ure{\~n}a-L{\'o}pez, L. Alfonso and Mart{\'i}nez C{\'a}mara, Eugenio and Camacho-Collados, Jose},
  booktitle = {Proceedings of the 2024 Joint International Conference on Computational Linguistics, Language Resources and Evaluation},
  pages     = {13471--13488},
  year      = {2024},
  url       = {https://aclanthology.org/2024.lrec-main.1179/}
}

@inproceedings{oses2025semeval,
  title     = {{SemEval}-2025 Task 8: Question Answering over Tabular Data},
  author    = {Os{\'e}s Grijalba, Jorge and Ure{\~n}a-L{\'o}pez, L. Alfonso and Mart{\'i}nez C{\'a}mara, Eugenio and Camacho-Collados, Jose},
  booktitle = {Proceedings of the 19th International Workshop on Semantic Evaluation},
  pages     = {2512--2522},
  year      = {2025},
  url       = {https://aclanthology.org/2025.semeval-1.324/}
}

@article{tian2024spreadsheetllm,
  title   = {{SpreadsheetLLM}: Encoding Spreadsheets for Large Language Models},
  author  = {Tian, Yuzhang and Zhao, Jianbo and Dong, Haoyu and Xiong, Junyu and Xia, Shiyu and Zhou, Mengyu and Lin, Yun and Cambronero, Jos{\'e} and He, Yeye and Han, Shi and Zhang, Dongmei},
  journal = {arXiv preprint arXiv:2407.09025},
  year    = {2024},
  url     = {https://arxiv.org/abs/2407.09025}
}

@inproceedings{li2025mimotable,
  title     = {{M}i{M}o{T}able: A Multi-scale Spreadsheet Benchmark with Meta Operations for Table Reasoning},
  author    = {Li, Zheng and Du, Yang and Zheng, Mao and Song, Mingyang},
  booktitle = {Proceedings of the 31st International Conference on Computational Linguistics},
  pages     = {2548--2560},
  publisher = {Association for Computational Linguistics},
  address   = {Abu Dhabi, UAE},
  year      = {2025},
  url       = {https://aclanthology.org/2025.coling-main.173/}
}

@article{pan2022trag,
  title   = {End-to-End Table Question Answering via Retrieval-Augmented Generation},
  author  = {Pan, Feifei and Canim, Mustafa and Glass, Michael and Gliozzo, Alfio and Hendler, James},
  journal = {arXiv preprint arXiv:2203.16714},
  year    = {2022},
  url     = {https://arxiv.org/abs/2203.16714}
}

@article{zou2025trag,
  title   = {{RAG} over Tables: Hierarchical Memory Index, Multi-Stage Retrieval, and Benchmarking},
  author  = {Zou, Jiaru and Fu, Dongqi and Chen, Sirui and He, Xinrui and Li, Zihao and Zhu, Yada and Han, Jiawei and He, Jingrui},
  journal = {arXiv preprint arXiv:2504.01346},
  year    = {2025},
  url     = {https://arxiv.org/abs/2504.01346}
}

@article{yu2025tablerag,
  title   = {{TableRAG}: A Retrieval Augmented Generation Framework for Heterogeneous Document Reasoning},
  author  = {Yu, Xiaohan and Jian, Pu and Chen, Chong},
  journal = {arXiv preprint arXiv:2506.10380},
  year    = {2025},
  url     = {https://arxiv.org/abs/2506.10380}
}

@article{gulati2026frtr,
  title   = {From Rows to Reasoning: A Retrieval-Augmented Multimodal Framework for Spreadsheet Understanding},
  author  = {Gulati, Anmol and Sen, Sahil and Sarguroh, Waqar and Paul, Kevin},
  journal = {arXiv preprint arXiv:2601.08741},
  year    = {2026},
  url     = {https://arxiv.org/abs/2601.08741}
}

@article{gulati2026brtr,
  title   = {Beyond Rows to Reasoning: Agentic Retrieval for Multimodal Spreadsheet Understanding and Editing},
  author  = {Gulati, Anmol and Sen, Sahil and Sarguroh, Waqar and Paul, Kevin},
  journal = {arXiv preprint arXiv:2603.06503},
  year    = {2026},
  url     = {https://arxiv.org/abs/2603.06503}
}

\appendix

\section{Operational Baseline and Method Details}
\label{app:method}

The prefix-RAG parser detects sheet boundaries, scores candidate header rows, and retains up to three sample rows per sheet. The resulting text is truncated to 2{,}000 tokens, embedded as one document-level unit, and retrieved as context for the generator. Because the full table is unavailable at answer time, global aggregation usually cannot be computed.

For the 2{,}668-row Forbes billionaires table, the question ``How many billionaires are under 40?'' is answered from three sample rows by the reproduction. \fico{} instead executes:

\begin{nolinenumbers}
\begin{lstlisting}[language=SQL]
SELECT COUNT(*) FROM data
WHERE age < 40 AND age IS NOT NULL;
-- Result: 87
\end{lstlisting}
\end{nolinenumbers}

\paragraph{Representative document summary.}
\begin{quote}
\small\itshape
``This dataset tracks 2{,}668 billionaires from the Forbes World's Billionaires list. It covers net worth, age, self-made status, gender, philanthropy score, and country of origin. Key columns: personName, finalWorth, age, selfMade, gender, philanthropyScore, country. It can answer questions about wealth, country distributions, age breakdowns, and wealth-source categories.''
\end{quote}

\paragraph{Representative execution prompt.}\mbox{}\par\vspace{-0.8em}
\begin{nolinenumbers}
\begin{lstlisting}
Table: data (2,668 rows)
Columns:
  personName: text, e.g. Elon Musk, Jeff Bezos
  finalWorth: numeric, range 1,000 to 219,000
  age: numeric, range 19 to 96
  selfMade: boolean, e.g. True, False
  gender: text, e.g. M, F
  country: text, e.g. United States, China
Sample rows:
  Elon Musk | 219000 | 50 | True | M | United States
  Jeff Bezos | 171000 | 58 | True | M | United States
\end{lstlisting}
\end{nolinenumbers}

\section{Controlled Evaluation Details}
\label{app:controlled}

The component runner changes one factor at a time over the same DataBench questions. The gold-workbook condition does not reveal the gold answer and retains sheet selection. Denotation is evaluated on a canonical result object before free-form synthesis. All questions remain in the denominator, including parse failures, safety rejections, and execution errors.

Removing summaries replaces the language-model summary with deterministic schema text. Disabling reranking preserves summary retrieval but uses its original candidate order. Disabling fallback prevents a move to another candidate after an empty result. The value-grounding ablation removes both representative column values and the three sample rows while retaining column names, inferred types, row counts, and generation rules.

The matched source-oracle experiment reuses cached outputs for the 1{,}472 questions on which \fico{} selected the gold workbook and regenerates only the 338 mismatched-source cases. Cache identity includes question, workbook, and compute configuration. We confirmed equality of SQL, candidate traces, verifier decisions, denotations, and correctness for every shared-workbook pair. The gold-minus-selected contingency table is: 1{,}191 both correct, 420 both wrong, 190 gold only, and 9 selected only.

\begin{table}[H]
\centering
\small
\setlength{\tabcolsep}{4pt}
\begin{tabular}{lrrr}
\toprule
\multicolumn{4}{l}{\textit{A. Find stage: DataBench end to end}} \\
\textbf{Variant} & \textbf{Doc rank-1} & \textbf{Correct doc} & \textbf{Final} \\
\midrule
Controlled full & 65.4\% & 65.3\% & 21.3\% \\
$-$ summaries (schema only) & 61.9\% & 61.6\% & 18.7\% \\
$-$ schema reranking & 65.4\% & 65.1\% & 20.4\% \\
$-$ empty-result fallback & 65.4\% & 65.7\% & 21.3\% \\
\midrule
\multicolumn{4}{l}{\textit{B. Compute stage: DataBench gold-workbook condition}} \\
\textbf{Variant} & \textbf{Executes} & \textbf{Denotation} & \textbf{Final} \\
\midrule
Full schema prompt & 72.3\% & 25.9\% & 27.8\% \\
$-$ representative/sample values & 72.3\% & 18.3\% & 20.8\% \\
\bottomrule
\end{tabular}
\caption{The controlled interventions isolate two useful mechanisms: summaries improve final accuracy by 2.6 points, while representative values and sample rows improve it by 7.0. Absolute scores are not compared across runners.}
\label{tab:ablations}
\end{table}

\section{Representative End-to-End Cases}
\label{app:cases}

\begin{table}[H]
\centering
\small
\begin{tabularx}{\textwidth}{YYY}
\toprule
\textbf{Question} & \textbf{Prefix-RAG reproduction} & \textbf{\fico{}} \\
\midrule
``Among the top five countries by reported coronavirus cases, which has the highest case-fatality ratio?'' & ``I don't know. The context only lists the United States.'' & Brazil (2.189\%) from a grouped ratio query. \\
\midrule
``Among non-drinkers, is gambling mostly Yes or No?'' & ``Based on context, I cannot determine.'' & Mostly No: 89 No versus 33 Yes. \\
\midrule
``Do PASS or FAIL movies have a higher profitable share?'' & ``I don't know. The context does not contain financial data.'' & PASS: 64.2\% versus FAIL: 61.8\%. \\
\bottomrule
\end{tabularx}
\caption{Representative FiCo wins: full-table execution answers global aggregations whose required rows are absent from the retrieved prefix.}
\end{table}

\section{Additional Results}

\begin{table}[H]
\centering
\small
\begin{tabular}{lrrr}
\toprule
\textbf{Answer type} & \textbf{Questions} & \textbf{Prefix-RAG} & \textbf{\fico{}} \\
\midrule
Boolean & 387 & 18.1\% & 86.6\% \\
Number & 412 & 10.2\% & 73.3\% \\
Category & 333 & 17.7\% & 80.2\% \\
List[category] & 329 & 10.3\% & 69.3\% \\
List[number] & 349 & 8.0\% & 70.8\% \\
\bottomrule
\end{tabular}
\caption{FiCo improves over prefix RAG for every DataBench answer type; gains range from 59.0 to 68.5 points.}
\end{table}

The residual audit assigns 32\% of the 431 joint misses or regressions to column confusion, often involving uninformative names such as \texttt{Unnamed\_2}. Near-duplicate workbook selection accounts for 31\%, benchmark or answer-format noise for 22\%, and wide-format tables for 15\%. These categories are descriptive because independent re-annotation and inter-annotator agreement are not reported.

\section{Additional Limitations}

The local parser approximates a deployed system, and implementation differences can change baseline accuracy. MiMoTable reference answers may not match terse database outputs, so human adjudication is needed before treating its score as a lower bound. The evaluation does not establish production-grade SQL safety or an enterprise data boundary for summaries and sample values.

\end{document}